\documentstyle[12pt]{article}
\title{ardo}
\author{Ardo}
\begin{document}                
\baselineskip=23pt
\parindent=20pt
\textfraction{0.1}
\topfraction{0.9}
\bottomfraction{0.9}
\title{\hspace{1 cm} GROWTH KINETICS IN THE
$\Phi ^6$ N-COMPONENT MODEL. Conserved Order Parameter}

\author{F.Corberi$^{1}$ and U.Marini.Bettolo Marconi$^{2}$}
\maketitle

\noindent
\newline
$1$ Dipartimento di Scienze Fisiche, Universit\'a degli studi di Napoli,
Mostra d'Oltremare, Padiglione 19, 80125 Naples, Italy
\newline
$2$ Dipartimento di Matematica e Fisica, Universit\'a di Camerino, via Madonna
delle Carceri, 62032 Camerino, Italy
\date{}
\medskip
\vspace{1 cm}

\begin{abstract}

We extend the discussion of the growth kinetics
of the large-N time-dependent Ginzburg-Landau model with an order
parameter described by a $\Phi^6$ free energy functional,
to the conserved case.
Quenches from a high temperature initial state to a zero temperature
state are studied for different selections of parameters entering the
effective potential. In all cases we obtain the asymptotic form
of the structure factor. As expected in analogy with the well studied $\Phi^4$
model, we find multiscaling behavior whenever stable equilibrium is reached
in the ordering region. On the other hand
the present model also displays  a novel feature, namely the
occurrence of metastable relaxation.

\end{abstract}
\newcommand {\be} {\begin{equation}}
\newcommand {\ee} {\end{equation}}
\setcounter{page}{1}

{\Large {\bf Introduction}}

In a previous paper~\cite{nostro}, hereafter referred as {\bf I},  we examined
the growth kinetics
of an N-component field with a non-linear local
interaction of the sixth order, namely a $\Phi^6$ term,
subject to a sudden quench from a high temperature
initial state. The equations of the large $N$
spherical model have
been deduced and solved in the case of a non-conserved order parameter (NCOP or
model A) at zero temperature.

In the present paper we turn our attention to the kinetics of
the  conserved parameter case (COP or model B).

In contrast with the non conserved case one can observe
multiscaling~\cite{mult}
behaviour whenever stable equilibrium configurations are reached in the
presence
of a potential with two degenerate minima (see {\bf I}, fig.1). The breakdown
of standard scaling~\cite{sca} is due to the same mechanism as for the $\Phi^4$
model~\cite{crz}: two distinct dominant lengths compete in the late stage of
the dynamical
process, namely the dimensional length $L(t)=(2 \Gamma t)^{1/z}$,
with $z=4$, and the
wavelength $k_m^{-1}$ of the peak of the structure factor. Since these two
lengths differ only by a logarithmic factor one cannot speak of a single
dominant length. As a consequence, each mode $k$ evolves with a different
exponent, a behaviour which has been named multiscaling.

 In analogy with the NCOP, the $\Phi ^6$ model exhibits also metastable
solutions which
are absent in the spherical $\Phi ^4$.
 In the following section, in fact, we
shall extend a theorem concerning the metastability of the model
which is the equivalent of the one already proved for the NCOP. This
suggests that the metastability is not due to the kind of relaxation
dynamics assumed, but it is a property of the functional form of the free
energy.

 The presence of metastability is interesting because, as stressed in {\bf I},
it is commonly believed that the metastable solutions disappear when taking the
limit $N\to \infty$.

The present paper is organized as follows:

In section $1$ we solve the model by explicit integration of the equations
of motion for the structure factor, in the asymptotic regime. Since these
equations have been obtained in the first part of this work we refer to
{\bf I}  for a more detailed presentation of the model and of its large-$N$
limit, as well as for a discussion of the initial conditions.
Finally
in the last part we summarize, discuss some result and conclude the work.

\section  {Solution of COP the model}

We start by considering the conserved model ($p=2$) explicitly and
solving the equation of motion for the structure factor, eq.(12) of {\bf I}.
In the present  case, introducing the scaling function
$F(\vec x)=\frac {e^{-x^4}}{x^\theta}$ of the
dimensionless wave-vector $\vec x=\vec k L(t)$, eq. (10) of {\bf I} can be
formally
integrated, yielding:
\be
C(\vec k,t)=\Delta e^{-\beta (t)x^2}L^\theta (t)\,F(\vec x)
\label{13}
\ee
where:
\be
\beta (t)=sign(Q(t)) \left [ \frac {{\cal L}(t)}{L(t)} \right ]^2
\label{14}
\ee
is the squared rate between the two lengths
${\cal L }(t)=[{2\Gamma |Q(t)|}]^{1/2}$ and $L(t)$,
and $Q(t)$ is obtained by self-consistency eliminating $S(t)$ from eq.(9)
with the help of eq.(10) of {\bf I}.  From this equation we obtain:
\be
S(t)=\Delta L(t)^{\theta-d}\int d^d x \, \, x^{-\theta} e^{-\beta (t)x^2 -x^4}
\label{16}
\ee
In order to calculate the integral on the right hand side one has to consider
three different cases, according to the behaviour of $\beta (t)$ in
the late stages~\cite{crz}:

\vspace{7 mm}

a) $\lim_{t\to \infty}\beta (t)=0$

\vspace{7 mm}

 In this case $L(t)$ prevails over ${\cal L}(t)$.
To first order in $\beta (t)$, one obtains asymptotically:
\be
S(t)\simeq \Delta L(t)^{\theta -d} \int d^d x x^{-\theta}[1-\beta (t)x^2]
e^{-x^4}=\Delta L(t)^{\theta -d} [I_o -\beta (t)I_1 ]
\label{17}
\ee
where:
\be
I_o=\int d^d x\, \, x^{-\theta} e^{-x^4}
\label{18}
\ee
and:
\be
I_1 =\int d^d x \, \,x^{2-\theta} e^{-x^4}
\label{19}
\ee

\vspace {7 mm}

b) $\lim_{t\to \infty} \beta (t) =\infty$

\vspace{7 mm}
 Now ${\cal L}(t)$ is asymptotically dominant.
Since the term $\beta (t)x^2$ prevails over $x^4$ in the exponential of
eq.~(\ref{16}), one can disregard the latter and obtain:
\be
S(t)\simeq \Delta L(t)^{\theta -d} \int d^d x\, \,
x^{-\theta} e^{-\beta (t)x^2}
\sim \Delta L(t)^{\theta -d}\beta (t)^{\frac {\theta -d}{2}}
\label{20}
\ee
in the late stages of the dynamics.

Finally one considers:

\vspace{7 mm}

c) $\lim _{t\to \infty}\beta (t)=-\infty$

\vspace{7 mm}

In this case $S(t)$ can be evaluated asymptotically by the steepest descent
technique:
\be
S(t)\sim  L(t)^{\theta -d}\Delta e^{\frac {\beta ^2 (t)}{4}}
\beta (t)^{\frac {\theta -d}{2}}
\label{21}
\ee
Because of this threefold possibility in the calculation of $S(t)$ we need
to know the qualitative behaviour of $S(t)$ in the late time regime. As in the
non conserved case [1],
it is sometimes (e.g. in cases $\mu_3$ and $\mu _6$) not a priori evident
whether the relaxation into the metastable $\Phi \equiv 0$ configuration is
to be expected. Here we are able, as for NCOP, to produce a criterion which
establishes under which conditions
metastable solutions are allowed.

A necessary and sufficient condition in order to observe relaxation in the
$\Phi \equiv 0$ final state (i.e. $\lim_{t\to \infty}S(t)=0$), is to have:
\be
\frac {\partial ^2 V(\phi)}
{\partial \phi ^2}|_{\phi =0} \geq 0
\label {crit1}
\ee
and:
\be
S(\overline t)\leq \frac{-g-\sqrt{g^2-4\lambda r}}{2\lambda})
\label {crit2}
\ee
where $\overline t$ is a generic time instant
in the asymptotic region.

We outline the proof of the criterion beginning with the necessary condition
eq.~(\ref{crit1}), i.e:
\be
\{ \lim_{t\to \infty}S(t)=0 \} \Rightarrow \{\frac {\partial ^2 V(\Phi)}
{\partial \Phi ^2} |_{\Phi =0} \geq 0 \}
\ee
In fact, since:
\be
\frac {\partial ^2 V(\phi )}{\partial \phi ^2}|_{\phi =0}=r
\ee

in the case $r<0$, recalling that $\lim_{t\to \infty }S(t)=0 $,
eq.(9) of {\bf I} reads:
\be
\dot Q(t)\simeq r
\ee
for long times. From eq.~(\ref{14}) one sees that $\lim_{t\to \infty} \beta (t)
=-\infty$. Inserting this result into eq.~(\ref{21}), we obtain:
\be
S(t)\sim e^{\frac {r^2 L^4(t)}{4}}L(t)^{2(\theta -d)}
\ee
which is not consistent with the assumption $\lim_{t\to \infty}S(t)=0$.
We conclude that it must be $r\geq 0$.
The explicit solution of the model proves
that consistent solutions effectively exist in this case (see section 2).

Secondly, as regards sufficiency, both eqs.~(\ref{crit1}) and ~(\ref{crit2})
are required simultaneously. In fact, from eq.(9) of {\bf I} we deduce that, if
the condition~(\ref{crit2}) is fulfilled, then:
\be
\dot Q(\overline t)\geq 0
\label{app1}
\ee
We distinguish three asymptotic scenarios:

\vspace {7 mm}

a)$\lim_{t\to \infty}\beta (t)=0$.

\vspace{7 mm}

In this case, from eq.~(\ref{17}),
$\lim_{t\to \infty }S(t)=0$. [q.e.d.]

\vspace{7 mm}

b)$\lim_{t\to \infty}\beta (t)=\infty$.

\vspace{7 mm}

Now, from eq.~(\ref{20}), we have:
\be
\frac {\dot S(t)}{S(t)}=-(\frac {d-\theta }{2})\frac {\dot Q(t)}{Q(t)}
\label{app2}
\ee
In this case, since $S(t)$ is positive defined,
from eq.~(\ref{app1}) and eq.~(\ref{app2}) we deduce that $\dot S(\overline
t)<0$
and therefore $\lim_{t\to \infty}S(t)=0$. [q.e.d.]

\vspace{7 mm}

c)$\lim_{t\to \infty}\beta (t)=-\infty$.

\vspace{7 mm}

In this case, from eq.~(\ref{21})
we get:
\be
\frac {\dot S(t)}{S(t)}=(\frac {d-\theta}{2})\frac {\dot Q(t)}{|Q(t)|}
[-\frac {2\Gamma Q^2 (t)}{d-\theta}(1+\frac {1}{2t} \frac {|Q(t)|}{\dot Q(t)})
+1]
\label{app3}
\ee
In this case, since $\lim_{t\to \infty }Q(t)=-\infty $
from eq.~(\ref{app1}) and eq.~(\ref{app3}) we deduce that $\dot S(\overline
t)<0$
and so  $\lim_{t\to \infty}S(t)=0$. [q.e.d]

To summarize we observe that also in this case, as for NCOP, the possibility
of metastable relaxation, established by this criterion, is due to a local
property of the functional form of the potential around the metastable
solution,
i.e. eq.~(\ref{crit1}), and to a dynamical property of the asymptotic regime,
i.e. eq.~(\ref{crit2}). In practice a numerical solution of eq. (12) of {\bf I}
reveals that, when eq.~(\ref{crit1}) is fulfilled, it is always possible to
achieve the condition~(\ref{crit2}) by decreasing
the variance of the initial condition $S(0)$.

We turn now to the solution of the model by considering different cases,
according to the parameters $\mu \equiv
(r,g,\lambda)$ characterizing $V(\vec \Phi)$ (see {\bf I}, fig.1).

Let us begin with the case of simple diffusion:

\vspace {1 cm}

${\bf \mu _o \equiv (r=0,g=0,\lambda=0)}$

\vspace {1 cm}

This case is trivial to compute since $Q(t)\equiv 0$, but interesting because
the existence of a fixed point at $\mu _0$ affects the behaviour of dynamical
processes characterized by different values of the parameters $\mu$. By
explicitly calculating the structure factor, from eq.~(\ref{13}) we find:
\be
C(\vec k,t)=\Delta L^\theta (t) F(\vec x)
\label{22}
\ee
This result, which is exact at all times (and for all $N$), shows that scaling
holds true from
beginning to end, as in the non-conserved case.

\vspace{1 cm}

${\bf \mu _1 \equiv (r=0,g=0,\lambda >0)}$

\vspace {1 cm}

This choice of the parameters represent the tricritical~\cite{tric} case which
can
be solved by considering, firstly, the case
a) $\lim_{t\to \infty}\beta (t)=0$.
{}From eq.(9) of {\bf I} and eq.~(\ref{17}), to leading order, one obtains:
\be
Q(t)\simeq aL(t)^{2(\theta -d+2)}+c
\label{23}
\ee
where $a=\frac {\lambda \Delta ^2 I_o ^2}{\Gamma (\theta -d+2)}$ and c
are constants.

{}From eq.~(\ref{23}) we evaluate:
\be
\beta (t)\sim L(t)^{2(\theta -d+1)}+cL^{-2}(t)
\label {24}
\ee
Therefore this solution is consistent only for $d>\tilde d_c$, with
$\tilde d_c=\theta +1$, as for NCOP. In this case we find:
\be
C(\vec k,t) \simeq  \Delta e^{-[aL(t)^{2(\tilde d_c -d)}+cL(t)^{-2}]x^2}
L^\theta (t)\, F(x)
\label{25}
\ee
Hence:
\be
C(\vec k,t)\simeq \Delta [1-aL(t)^{2(\tilde d_c -d)}x^2 ]
L^\theta (t)\, F(\vec x)
\label{26}
\ee
for $\tilde d_c <d<\tilde d_c +2$, while:
\be
C(\vec k,t)\simeq \Delta [1-\tilde c L(t)^{-2}x^2]L^\theta (t)\, F(\vec x)
\label{27}
\ee
for $d\geq \tilde d_c +2$ (where $\tilde c=c$ for $d>\tilde d_c +2$).

So, for $d>\tilde d_c$, the trivial fixed point at $\mu_0$ is still
attractive and the system scales asymptotically as in the
$\mu _o$ case, with $x$ dependent corrections to scaling.

We consider now quenches at $d\leq \tilde d_c$. In this case $\lim_{t\to
\infty}\beta (t)=\infty$ since $\beta (t)\to -\infty$ is not allowed for
$\mu _1$. From eq.(9) of {\bf I} and eq.~(\ref{20}) we deduce:
\be
Q(t)\simeq [(d-\theta +1)(aL^4(t)+c)]^{\frac{1}{d-\tilde d_c +2}}
\label{28}
\ee
with $a=\frac {\lambda \Delta ^2}{(2\Gamma)^{d-\tilde d_c +2}}$ and
$c$ is a constant. Therefore:
\be
\beta (t)\sim L(t)^{2 \frac {\tilde d_c -d} {d- \tilde d_c+2}}
\label{29}
\ee
which is consistent for $d<\tilde d_c$.
 In this case ${\cal L}(t)$ prevails asymptotically over $L(t)$ and we have
to look for scaling with respect to the former length. Therefore we go back to
eq. (1), which can be written as:
\be
C(\vec k,t)= \Delta e^{-k^4L^4(t)}{\cal L}^{\theta}(t) \tilde F(\tilde x)
\label{aaaa}
\ee
where the scaling function is now defined as:
\be
\tilde F(\tilde x)=\frac {e^{-\tilde x^2}}{\tilde x^\theta}
\label{bbbb}
\ee
and the dimensionless variable $\tilde x= k{\cal L}(t)$ is expressed in
terms of the dominant length ${\cal L}(t)$. Substituting, with the help of eq.
(\ref {29}), $L(t)$ with ${\cal L}(t)$ in eq. (\ref {aaaa}) we obtain:
\begin{eqnarray}
C(\vec k,t) &=& \Delta e^{-a{\cal L}^{2(d-\tilde d_c)\tilde x^4}}
{\cal L}^{\theta}(t) \tilde F(\tilde x) \simeq \nonumber \\
            &\simeq& \Delta \left
[1-a{\cal L}^{2(d-\tilde d_c)\tilde x^4} \right ] {\cal L}^{\theta}(t)
\tilde F(\tilde x)
\end{eqnarray}
where $a$ is a constant. This result shows that scaling holds true even in
this case, with $\tilde x$ dependent corrections, but with a modified
scaling function. The power growth law of the dominant length,
${\cal L}(t) \sim t^{\frac {1}{\tilde z}}$, is obeyed with an exponent
$\tilde z=2(d-\tilde d_c+2)$, which depends on the space dimensionality.

\vspace {1 cm}

${\bf \mu _2 \equiv (r=0,g>0,\lambda >0)}$

\vspace {1 cm}

In this case, as for NCOP, we expect the parameter $\lambda $ to be irrelevant
and the asymptotic form of the structure factor to be the same as for a
$\Phi ^4$ theory (with $r=0$) (see ~\cite{crz}).
In fact, since $\lim_{t\to \infty} S(t)=0$,
solving eq. (9) of {\bf I} for long times:
\be
\dot Q(t)\simeq g\, S(t)
\label{31}
\ee
we find:
\be
C(\vec k,t)\sim \Delta e^{-g\Delta a' L(t)^{\frac {2(d_c -d)}{d-d_c +4}}x^2}
L^\theta (t)\, F(\vec x)
\label{34a}
\ee
for $d<d_c=\theta +2$
\be
C(\vec k,t)\sim \Delta [1-g\Delta a L(t)^{d_c -d}x^2]L^\theta (t)\, F(\vec x)
\label{32a}
\ee
for $d_c <d<d_c +2$, and:
\be
C(\vec k,t)\sim \Delta [1-cL(t)^{-2}x^2]L^\theta (t)\, F(\vec x)
\label{33a}
\ee
when $d\geq d_c +2$. In eqs.~(\ref{34a}) ~(\ref{32a}) and ~(\ref{33a}) $a$,
$a'$ and
$c$ are constants and $d_c=\theta +2$ is a critical dimensionality playing a
role similar to that of $\tilde d _c$ in the quench at $\mu _1$.

\vspace {1 cm}

${\bf \mu _3 \equiv (r=0,g<0,\lambda >0)}$

\vspace {1 cm}

As in the NCOP model, the quench at $\mu _3$ is very peculiar because
the necessary condition ~(\ref{crit1}) suggests that
metastable relaxation to $\vec \Phi (\vec r,t=\infty )
\equiv 0$ is not ruled out, but the sufficient condition ~(\ref{crit2})
does not apply
here since it reduces to $S(\overline t) \leq 0$ which is never true for
finite times. We will show that both the dynamics
leading to stable and to metastable equilibrium
are consistent with the model equations.

 When stable equilibrium is reached, since in this case $\lim_{t\to \infty}
S(t)=-\frac {g}{\lambda }\neq 0$ from eqs.~(\ref{17}), ~(\ref{20}) and
{}~(\ref{21}) we observe that only
$\lim_{t\to \infty}\beta (t)=-\infty $ is possible.
 Therefore, from eq.~(\ref{21}), letting $S(t)\simeq -\frac {g}{\lambda}$,
it is found:
\be
Q(t)\sim -(\frac {d-\theta}{2\Gamma }\, t \, \, ln\,t)^{\frac {1}{2}}
\label{35a}
\ee
and eventually:
\be
C(\vec k,t)\sim \Delta e^{[(d-\theta )ln\, t]^{\frac {1}{2}}x^2}L^\theta (t)\,
F(\vec x)
\label{36a}
\ee
As in the $\Phi^4$ theory standard scaling is broken in the limit of
large $N$. Instead, a multiscaling symmetry shows up in eq.~(\ref{36a}).
By considering the peak of the structure factor $k_m(t)$
it has been found:
\be
[k_m (t)L(t)]^4\simeq \frac {d-\theta }{4} ln\, t
\label {37}
\ee
In other words
in the late stages of the quench, two distinct lengths $k_m^{-1}$ and
$L(t)$ exist and  differ only by a logarithmic factor. The standard
scaling symmetry is broken by this
feature and, instead, multiscaling holds ~\cite{mult}.

 If metastable equilibrium is approached, however, assuming $\lim_{t\to \infty}
\beta (t)=0$, from eq.~(\ref{17}) and eq.(9) of {\bf I}, to leading order, we
find:
\be
Q(t)\simeq \frac {g\Delta a}{2\Gamma} L(t)^{\theta -d+4}+\tilde c
\label{38}
\ee
where $a=\frac {I_o}{\Gamma (\theta-d+4)}$ and $\tilde c$ are
constants.

Computing the ratio
\be
\beta (t)\sim L(t)^{\theta -d+2} +cL(t)^{-2}
\label{39}
\ee
we deduce that this solution is consistent only for $d>d_c =\theta +2$.
In this case we find:
\be
C(\vec k,t) \simeq \Delta e^{[g\Delta aL(t)^{d_c -d}+cL(t)^{-2}]x^2}
L^\theta (t)\,
F(\vec x)
\label{40}
\ee
Hence:
\be
C(\vec k,t)\simeq \Delta [1-g\Delta aL(t)^{d_c -d} x^2]L^\theta (t)\, F(\vec x)
\label {31}
\ee
for $d_c <d<d_c+2$, and:
\be
C(\vec k,t)\simeq \Delta [1-bL(t)^{-2}x^2]L^\theta (t)\, F(\vec x)
\label{32}
\ee
for $d\leq d_c +2$ (with $b=c$ for $d<d_c+2$ and $b=a+c$ for $d=d_c+2$).

Metastable equilibrium is approached with the same asymptotic dynamics
as in a quench at $\mu _2 $, for $d>d_c$.

On the other hand, for $d<d_c$, none of the asymptotic forms ~(\ref{20})
and ~(\ref{21}) are consistent. In fact let us try firstly with~(\ref{20}):
\be
S(t)\sim Q^{\frac {\theta -d}{2}}
\label{33}
\ee
If metastability is approached:
\be
\dot Q(t)\sim -|g|Q^{\frac {\theta -d}{2}}
\label {34}
\ee
in the late stage.
Since:
\be
\lim _{t\to \infty}Q(t)=+\infty
\label{35}
\ee
from eq.~(\ref{34}) it is:
\be
\lim _{t\to \infty }\dot Q(t)=-\infty
\label {36}
\ee
Statements ~(\ref{35}) and ~(\ref{36}) cannot be true simultaneously.

On the other hand, even the asymptotic form ~(\ref{21}) can never be consistent
with the requirement $S(t)\to 0$. In this case, in fact, eq.( 9) of {\bf I}
would read:
\be
\dot Q(t)=-|g|S(t)\simeq -|g||Q(t)|^{\frac {\theta -d}{2}}e^{\frac
{\Gamma Q^2(t)}{2t}}
\label {ppoi}
\ee
asymptotically. This equation, however, quickly leads to a diverging
$\dot Q(t)$,
which is not consistent with metastability (i.e. $S(t)\to 0$).

To summarize this $\mu _3$ case we observe that the situation is not
qualitatively
different from the NCOP case, in that for $d\leq d_c$ stable equilibrium
is always achieved, while, for $d>d_c$,
both stable and metastable solutions are possible and one passes from the
former to the latter by changing $\Delta$, the variance of the initial
condition.

\vspace {1 cm}

${\bf \mu _4 \equiv (r<0,g,\lambda >0)}$

\vspace {1 cm}

In this case statement ~(\ref{crit1}) prevents metastability. Therefore only
$\lim_{t\to \infty }\beta (t)=-\infty$ is consistent and, from eq.~(\ref{21}),
letting:
\be
S(t)\simeq \frac {-g+\sqrt {g^2 -4\lambda r}}{2\lambda}
\ee
for long times, the same form ~(\ref{35a}) is found for $Q(t)$, as in the
previous case. Hence:
\be
C(\vec k,t)\sim \Delta e^{[(d-\theta)ln\, t]^{\frac {1}{2}}x^2}L^\theta (t)\,
F(\vec x)
\label{48}
\ee
as for stable relaxation at $\mu _3$.

\vspace {1 cm}

${\bf \mu _5 \equiv (r>0,g\geq -4\sqrt {\frac {\lambda r}{3}},\lambda >0)}$

\vspace {1 cm}

Now $\lim_{t\to \infty}S(t)=0$. Hence, solving eq. (9) of {\bf I} for
long times
we find:
\be
Q(t)\simeq rt+c
\label{49}
\ee
where $c$ is a constant. Therefore ${\cal L}(t) \sim \sqrt{r} L^2(t)$
is the dominant length and, by means of eq. (\ref{aaaa}) we obtain:
\begin{eqnarray}
C(\vec k,t) &=& \Delta e^{-\frac{ \tilde x^4}{r {\cal L}^2(t)}}
\tilde F(\tilde x) \simeq \nonumber \\
            &\simeq& \Delta \left
[1-\frac{ \tilde x^4}{r {\cal L}^2(t)} \right ]
\tilde F(\tilde x)
\end{eqnarray}
This result shows that, differently from the corresponding NCOP case,
scaling holds controlled by the length ${\cal L}(t) $, which grows as
$t^{\frac{1}{\tilde z}}$, with $\tilde z=2$ in any dimension.

\vspace {1 cm}

${\bf \mu _6 \equiv (r>0,g<-4\sqrt {\frac {\lambda r}{3}},\lambda >0)}$

\vspace {1 cm}

According to statements ~(\ref{crit1}) and ~(\ref{crit2}) if the dynamics leads
asymptotically
the system to a
state with
$S(\overline t)\leq \frac {|g|-\sqrt {g^2 -4\lambda r}}{2 \lambda}$, metastable
relaxation in $\vec \Phi \equiv 0$ occurs. Practically
this is always achievable by choosing an initial condition with $S(0)$
sufficiently small. In this case the
equations can be solved asymptotically as in a quench at $\mu _5$.
The same results are obtained:
\be
C(\vec k,t) \sim \Delta \left
[1-\frac{ \tilde x^4}{r {\cal L}^2(t)} \right ]
\tilde F(\tilde x)
\label {51}
\ee
In this case, as compared to $\mu _3$, the stronger character of metastability
eliminates the critical dimension: metastable relaxation occurs in any
dimension when $S(0)$ is small. Since ~(\ref{crit2}) is only a sufficient
condition,
eq.~(\ref{51}) can in principle hold asymptotically even for $S(\overline t)$
larger than $\frac {|g|-\sqrt {g^2 -4\lambda r}}{2\lambda }$.

When $S(0)$ is sufficiently large, however, stable equilibrium must be
obtained. In this case, proceeding as for $\mu _3$ or $\mu _4$, the same
multiscaling
solution ~(\ref{36a}) is obtained in the late stages:
\be
C(\vec k,t)\sim \Delta e^{[(d-\theta)ln\, t]^{\frac {1}{2}}x^2}L^\theta (t)\,
F(\vec x)
\ee

\section {Summary}

In this paper we have extended the solution of the spherical $\Phi ^6$ model to
the conserved case (model B), with zero temperature.

To summarize the results of the present paper we observe that the $\Phi^6$
model shows either scaling or multiscaling in the asymptotic regime depending
on the
parameters $\mu$ of the Hamiltonian.
 The asymptotic multiscaling, peculiar to the large-$N$ limit, shows up,
as expected, in those cases when stable equilibrium is reached in a potential
with two degenerate minima (see {\bf I}, fig.1).
 The standard scaling behaviour, on the other hand, is controlled
either by $L(t)$ or by ${\cal L}(t)$, with different exponents and
scaling functions in the two cases.

The scaling behaviour controlled by $L(t)$ is always induced by the presence of
the trivial fixed
point of simple diffusion at $\mu _o$, which can be attractive on the whole
$r=0$ axis. For $g\geq 0$ this line represents the edge of the sector $\mu_5$
where ${\cal L}(t)$ prevails asymptotically. As a consequence the trivial
fixed point is attractive only above a dynamical critical dimension, while
below the structure factor scales with ${\cal L}(t)$
as at $\mu_5$.

The negative part of the $g$ axis, on the contrary, is the intersection set of
two sector ($\mu_4$ and $\mu_6$) where multiscaling holds. Therefore the
trivial fixed point (which represent now metastable relaxation) competes, now,
with the multiscaling fixed point and, again, it can be attractive only for
dimensions which exceed the critical one, in analogy with static critical
phenomena.

Finally, for what concerns the $\mu_6$ region, it behaves as a transition
sector between the $\mu_4$ and $\mu_5$ sectors of the phase diagram, in that
both multiscaling and scaling induced by ${\cal L}(t)$ are found
(the last leads to
metastable equilibrium). Since $r>0$ the trivial fixed
point is never attractive
and no critical dimension is found.

As regards metastability, it occurs in strict analogy with the non-conserved
case, apart from a greater technical complexity. In both cases, in fact, a
similar criterion links metastability to a local property (its curvature) of
the potential around the metastable extremum. We conclude that the kind of
dynamics (whether conserved or not) does not affect the occurrence of
metastability which is due, on the contrary, to the analytical properties of
the Hamiltonian.This is an interesting feature which suggest an unified
description of physical processes involving thermodynamic metastability.

\section {Acknowledgements}

We are grateful to A.Coniglio and M.Zannetti for many discussions about
scaling and multiscaling in spherical models.
\newpage

\section {Figure caption}

In figure 1 various shapes of the potential
$V(|\vec\phi|)$ are schematically shown
as a function of the parameters
$\mu\equiv [r,g,\lambda]$, for $\lambda>0$.

\end{document}